\documentclass[fleqn,10pt]{wlscirep}
\title{Current Induced Resistive State in Fe(Se,Te) Superconducting Nanostrips}

\author[1,*]{Ciro Nappi}
\author[1]{Carlo Camerlingo}
\author[2]{Emanuele Enrico}
\author[3]{Emilio Bellingeri}
\author[3]{Valeria Braccini}
\author[3]{Carlo Ferdeghini}
\author[1,+]{Ettore Sarnelli}
\affil[1]{CNR-SPIN, Sede secondaria di Napoli,  I-80078 Pozzuoli, Napoli (NA), Italy}
\affil[2]{INRIM, Istituto Nazionale di Ricerca Metrologica, I-10135 Torino, Italy} 
\affil[3]{CNR-SPIN, Genova, Corso Perrone 24, I-16152 Genova, Italy}
\affil[*]{ciro.nappi@spin.cnr.it}
\affil[+]{ettore.sarnelli@spin.cnr.it}

\begin{abstract}
We study the current-voltage characteristics of Fe(Se,Te) thin films deposited on CaF$_2$ substrates in form of nanostrips (width $w\sim\lambda$, $\lambda$ the London penetration length). In view of a possible application of these materials to superconductive electronics and micro-electronics we focus on transport properties in small magnetic field, the one generated by the bias current. From the characteristics taken at different temperatures we derive estimates for the pinning potential $U$ and the pinning potential range  $\delta$ for the magnetic flux lines (vortices). Since the sample lines are very narrow,  the classical creep flow model  provides a sufficiently accurate interpretation of the data only when the attractive interaction between magnetic flux lines of opposite sign  is taken into account. The observed voltages and the induced depression of the critical current of the nanostrips are compatible with the presence of a low number ($\lesssim 10$) magnetic field lines at the equilibrium, a strongly inhomogeneous current density distribution at the two ends of the strips and a reduced Bean Livingston barrier. In particular, we argue that the sharp corners defining the bridge geometry represent points of easy magnetic flux lines injection. The results are relevant for creep flow analysis in superconducting Fe(Se,Te) nanostrips. 
\end{abstract}
\begin{document}
\flushbottom
\maketitle
%
%
\thispagestyle{empty}

\section*{Introduction}
\begin{figure}[ht]
\centering
\includegraphics[width=15cm]{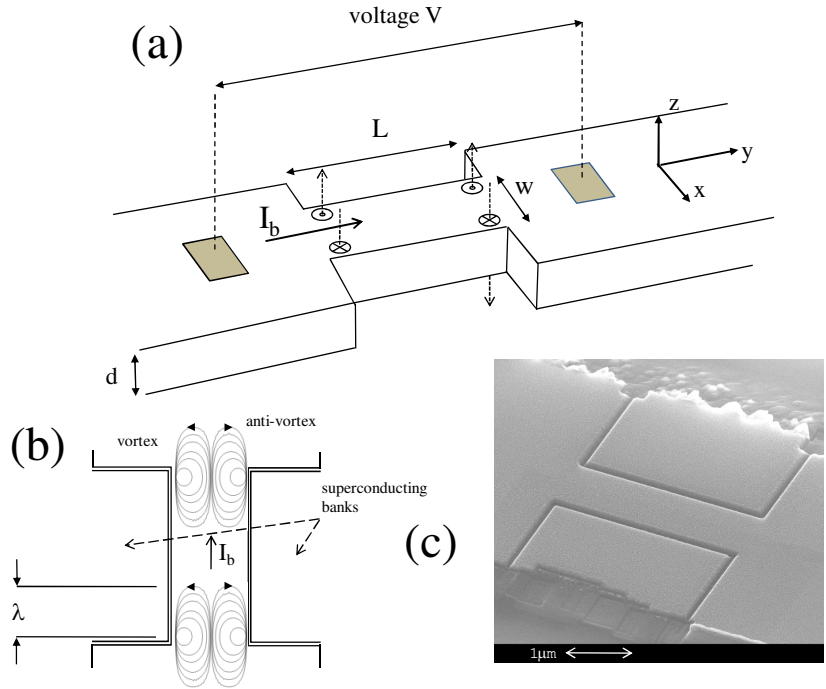}
\caption{(a) Geometry of the nanostrips used in this work. Also shown the self-generated  magnetic field lines, or vortices, (dashed arrows) entering the sample in the presence of a current bias $I_b$ and in correspondence of the four nanostrip corners.(b) Schematic representation of the  streamlines of the vortex and anti-vortex current densities before annihilation, the arrows shows the vortex current density direction. (c) A SEM image of sample B (w = 800 nm).}
\label{fig:micro}
\end{figure}

Currently, iron based superconductors are object of intense investigations as concerns their fundamental properties\cite{kami,pagl,hosono,maz,nappi1,nappi2,sarn,sarn1,sarn2,parl}. A potential use of these materials is expected both in the field of large scale current transport \cite{gao} and in micro-electronics or nano-electronics applications \cite{seid}. As is well known,  the presence of mobile magnetic flux lines in superconductor samples affects critically their current transport properties even when a magnetic field is not expressly applied \cite{tala}. The magnetic field self-generated by the bias current is able by itself to create magnetic vortices that, when in motion, induce  dissipation in the sample under test. According to the creep flow model \cite{kimanderson} the degree of  dissipation in  a superconducting  film,  for a fixed current density, depends on the vortex pinning potential $U$ and on the density of pinning sites. In the case of micro-electronics and nano-electronics applications based on the new superconducting materials, like iron based  pnictides and chalcogenides, the study of the pinning energy and of the current transport under condition of weak magnetic fields is of fundamental interest. In these kind of applications the magnetic field experienced by the films, typically patterned in the form of submicron strip-lines, is as low as few tens of gauss. From this point of view, these investigations are of primary relevance analogously to those carried out under conditions of high magnetic fields, when the research perspective is high power applications. \cite{gure,lei, bell,leo1,leo2} Moreover understanding  creep flow mechanisms in new superconducting materials remains a challenging task with unexpected fundamental implications \cite{eley}.  
In this work we have investigated the current induced resistive state of narrow (width $w \lesssim \lambda <1 \mu$m, where $\lambda$ is the London penetration depth) Fe(Se$_{0.5}$,Te$_{0.5}$) iron-chalcogenide nanostrips. The current-voltage characteristics have been measured at different temperatures $T<T_c$ ($T_c$ being the superconductor critical temperature), at low current values and in the absence of an externally applied magnetic field. A current induced resistive state is observed. From our analysis we infer that the sharp corners defining the two ends of the nanostrips are preferred points of entrance for self generated magnetic vortices. The experimental results can be explained on the basis of the presence of few vortices obeying a conventional flux depinning model in the presence of a reduced Bean Livingston barrier. A pinning energy of order of few tens of meV, and a pinning range of few nanometers are estimated, compatibly with linear defects extending along the film thickness.  Although the analysis carried out with conventional flux creep models in zero external field \cite{enpu1,enpu2}  qualitatively accounts for the observed features, a corrective term, proportional to film thickness $d$ normalized to the width $w$ of the strip ($ d/w$) had to be introduced for improving the accuracy of the pinning energy estimate. This correction stems from the very narrow width, of nano-metric order, of the samples considered in the tests. The  attractive Lorentz-like force exerted between  magnetic vortex lines of opposite sign entering the two close opposite edges (vortex/anti-vortex interaction) is not negligible, differently from the case of large width samples ($w \sim 10 d$ or larger) where the self-generated magnetic flux lines are sufficiently separated most of their life time while crossing the strip. In some experimental situation like the one here described, at low bias currents this force may be as intense as the Lorentz force. Nevertheless, the obtained pinning energy results lower than the one reported in literature for  Fe(Se,Te) micro-bridges in the presence of intense magnetic fields \cite{leo1}.  The paper is organized as follows. In the next section we present the experimental data and justify an interpretation in terms of creep flow. In the Discussion section we: (i) review the mechanism underlying the onset of resistance in a superconducting strip driven by the bias current; (ii) identify several possible issues related to the scaling from micro to nano-scale of the samples and focus particularly on the vortex anti-vortex interaction effect on creep; (iii) calculate the pinning potential $U(T)$ with and without the effect of this interaction;(iv) try to identify the type of defects; (v) draw the conclusions. 
\begin{table}[ht]
\centering
\begin{tabular}{cccccc}
Sample&\mbox{$L$($\mu$m)}&\mbox{$w$(nm)}&\mbox{$T_c$(K)}&\mbox{$j_c$[4.2 K](A/cm$^2$)}\\
\hline
A&\mbox{3}&\mbox{500}&\mbox{13}&\mbox{$3.2\times10^4$}\\
B&\mbox{3}&\mbox{800}&\mbox{13}&\mbox{$8.1\times10^4$}\\
\end{tabular}
\caption{\label{tab:table1} Parameters of the nanostrips.}
\end{table}

\section*{Results}

\subsection*{Current Voltage Characteristics}
\label{sec:CVC}
\begin{figure}[htbp]
\centering
\includegraphics[width=\linewidth]{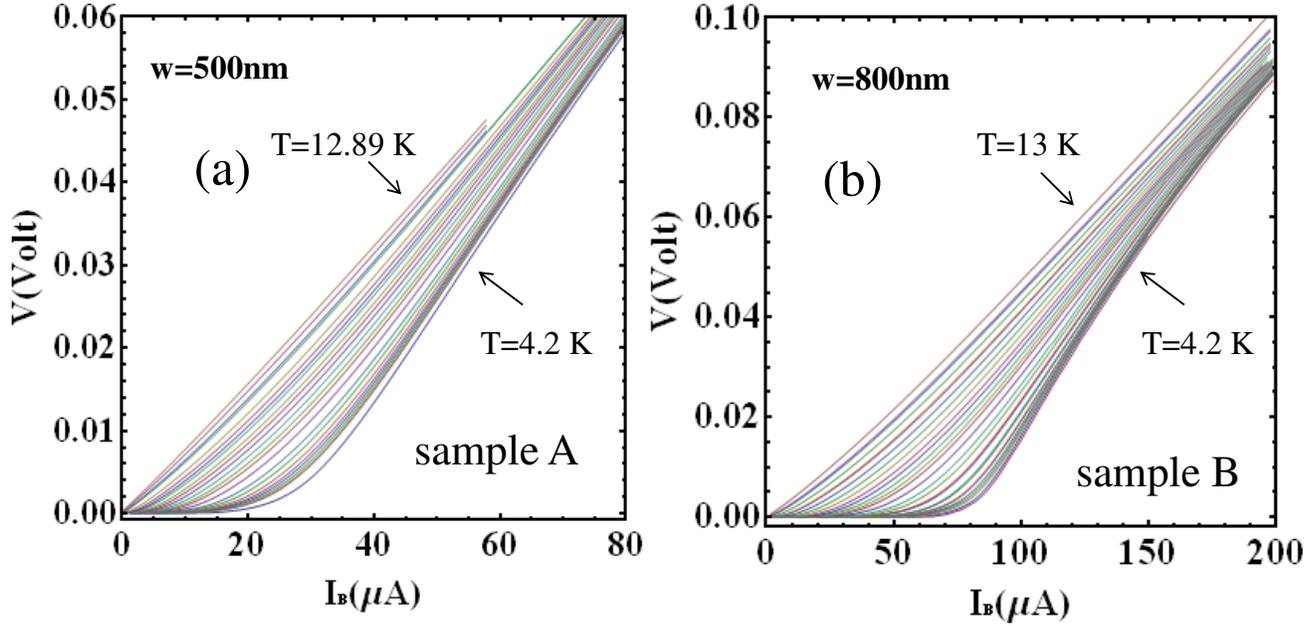}
\caption{Current voltage characteristics:(a) Sample A, w=500 nm. (b) Sample B, w=800 nm  for different  temperatures between 4.2 K and 13 K. Sample A: (T(K) = 4.2, 4.88, 4.97, 5.1, 5.29, 5.45, 5.72, 6.11, 6.38, 6.89, 7.31, 7.58, 7.88, 8.21, 8.5, 8.81, 9.19, 9.42, 9.74, 10.15, 10.96, 11.19, 11.84, 12.89). Sample B: (T(K) = 4.2, 4.61, 4.78, 5.0, 5.33, 5.42, 5.86, 5.99, 6.46, 6.54, 6.99, 7.35,7.83, 8.29, 8.53, 8.81, 9.12, 9.44, 9.82, 10.2, 10.62, 11.06, 11.13, 11.63, 12.16, 12.29, 13.0)}\label{fig:CVC}
\end{figure}

For the measurements of the current-voltage characteristics (CVCs), a Fe(Se$_{0.5}$,Te$_{0.5}$) film with thickness $d=100$ nm was patterned in the form of nanostrips with length $L=3$ $\mu$m and width $w=500$ nm (sample A) and $w=800$ nm (sample B), respectively. Fig. \ref{fig:micro} illustrates the geometry of our Fe(Se$_{0.5}$,Te$_{0.5}$)  nanostrips, while Table \ref{tab:table1} summarizes the experimental sample parameters. CVCs were collected at different temperatures between $4.2$ K and $12.89$ K for sample A (24 curves), and between $4.2$ K and $13$ K for sample B (27 curves). They are shown in  Fig. \ref{fig:CVC}.  As can be seen, the resistive state of CVCs emerges at finite temperatures, well below $T_c$, at small bias current densities, revealing the occurrence of creep flow, i.e. the vortex thermal depinning process. Creep flow, studied in literature chiefly in the presence of an external magnetic field \cite{gure,lei,bell,leo1,leo2,sarn}, is a manifestation  of the thermal agitation of the magnetic flux lines as they are acted upon, in the same time, by the adhesion force to crystalline defects and by the Lorentz-like force originated by the bias current. 
\begin{figure}[htbp]
\centering
\includegraphics[width=\linewidth]{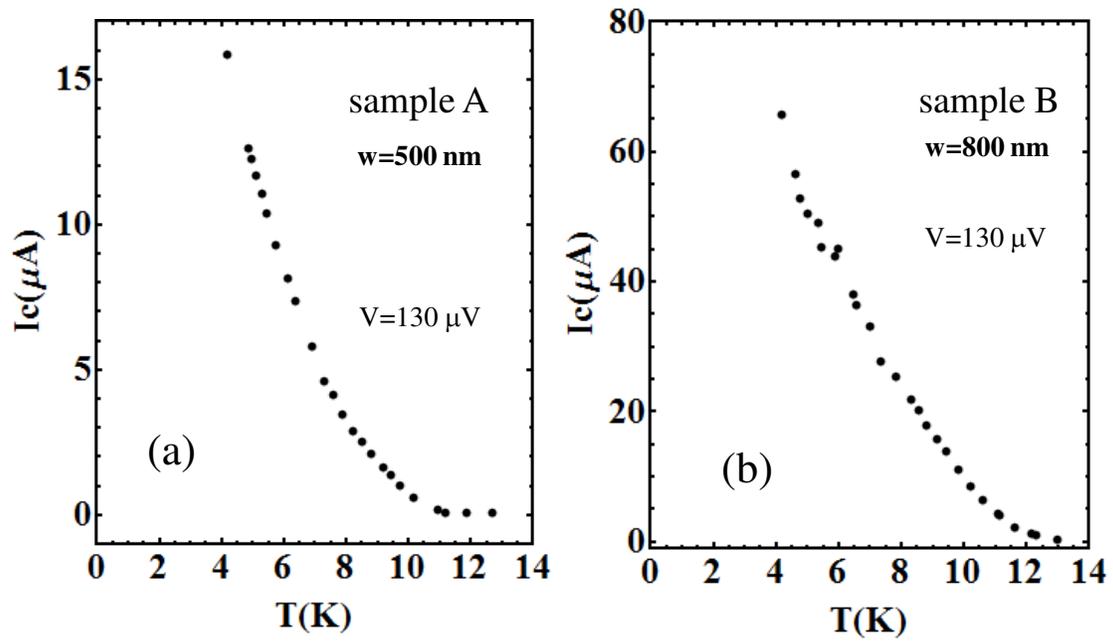}
 \caption{ Temperature dependence of the critical current $I_c$  (the current at which the voltage across the nanostrip overcomes the threshold $V=130$ $\mu$V) for (a), sample A and (b), sample B\label{fig:linear}}
\end{figure}
At higher bias current, when the effect of pinning becomes negligible, the Lorentz force generates a viscous motion of
quasi-free magnetic flux lines usually known as "flux flow".
Roughly speaking, the two phenomena manifest themselves in the CVC as strongly non linear and linear parts of the current-voltage plot, respectively. In this work we strictly focus on the non-linear part of the CVCs  observed at lower currents.
Among the parameters reported in Table \ref{tab:table1}, the critical current densities of the two samples, A and B at $T=4.2$ K appear, i.e. $j_c=3.2 \times 10^4$ and $j_c=8.1 \times 10^4$ A/cm$^2$ ($I_c=15.8 \mu$A, $I_c=65 \mu$A, see Fig. (\ref{fig:linear})), respectively; here, in the absence of an appreciable current threshold at $V=0$, and following a standard procedure, we have defined the critical current $I_c(T)$ as the current at which the voltage reaches the value $V=130$ $\mu$V across the Fe(Se$_{0.5}$,Te$_{0.5}$) nanostrips \cite{note}. The critical current densities were obtained by the ratio of the so defined critical current and the cross-sectional area of the strip ($S_A=5 \times 10^{-14}$m$^2$, $S_B=8 \times 10^{-14}$m$^2$). It should be noticed that  the slight reduction of $j_c$ found in sample A, as compared with sample B, is probably due to the detrimental of superconductivity properties caused by the etching process. In Fig. \ref{fig:linear}, the experimental dependence of the critical current on the temperature is shown. The critical current $I_c(T)$ decreases, far from $T_c$, almost linearly  with the temperature, which agrees with the interpretation  of our data in terms of creep flow \cite{mannh}. We point out that the extrapolated current density at $T=0$ of the data presented in Fig.\ref{fig:linear} gives the common value of $j_c(0)\sim 1.5 \times 10^5$ A/cm$^2$ for both samples. In our subsequent analysis, we will assume the following parameters for the superconductor: London penetration depth $\lambda(0)=560$ nm, coherence length $\xi(0)=2$ nm (Fe$_{1.03}$(Te$_ {0.63}$ Se$_{0.37})$ \cite{hyun}, an upper critical field $B_{c2}\sim 100$T \cite{gure,bell3}, in fair agreement with the theoretical estimate \cite{orl,tin} $B_{c2}=\phi_0/2 \pi \xi^2$=80 T.

\section*{Discussion}
\subsection*{Magnetic properties of the nanostrips}
The bias current injected in a superconducting strip (or wire) generates a magnetic self-field that limits the effective critical current density $j_c$ of the strip. Generally speaking, for a type II superconductor, the relevant critical field at which this dissipative condition emerges is the lower magnetic critical field: \cite{orl,tin,tala}
\begin{equation}
\label{eq:Hc1}
B_{c1}(T)=\mu_0 H_{c1}(T)=\frac{\phi_0}{4 \pi  \lambda(T)^2}\ln\left(\frac{\lambda}{\xi}\right)
\end{equation}
where $\lambda(T)/\xi(T)$ is the Ginzburg–Landau parameter $\kappa$, which is actually temperature independent, and $\mu_0$ is the vacuum magnetic permeability. At the field $H_{c1}$, magnetic field line penetration into the superconducting sample becomes energetically favourable.  As soon as the current in the nanostrips is sufficiently high, such that the magnetic field intensity  at the surface of the strip reaches $H_{c1}$, magnetic field lines of opposite sign will tend to symmetrically nucleate at the opposing edges and self-annihilate at the center of the nanostrips. The flux motion, in the same time, is strongly influenced by the tendency of the vortex normal cores (size $\sim \xi$) to stay pinned  to the crystalline defects of the material, a mechanism which lowers the free energy of the system and makes the vortex motion a hopping over the pinning sites. The pinning sites and the vortex-defect interaction are characterized by a pinning potential, or pinning energy, $U$, a hopping distance $\delta$ and a frequency of attempt $\omega_0$. In normal conditions,  this picture has to be refined because of the existence of a surface-flux line interaction effect, known as Bean Livingston barrier \cite{bean}. Even at $H_{c1}$, vortices may not enucleate from the edges and enter the film until  a stronger field $H_s$ (up to 20 times $H_{c1}$ and close to the thermodynamic critical field) is reached \cite{vino}.  At this field, the vortex attraction to the edge (the tendency of magnetic field lines to exit the bridge) is suppressed. Even though the average self field is considerably less than $H_s$, surface irregularities, defects, sample ends, proximity with further superconductors,  create local fields equal or greater to $H_s$. These irregularities constitute preferential points of ingress for the magnetic flux lines.
In our nanostrips two elements strongly dominate the magnetic behaviour: i) the sudden increase of the cross section at the two ends (presence of corners), where the supercurrent suddenly bents $90^\textrm{o}$ and ii) the presence of superconducting banks overlooking the nanostrips and separated by these latter by a gap of about $70$nm, (see SEM image in  Fig \ref{fig:micro}). As far as the first point is concerned, the presence of corners  produces the current crowding effect described by Clem \textit{et al.} \cite{clem} and it is  responsible for the observed early suppression of the superconducting state due to the vortex injection at these points;  regarding the second point, the two superconducting banks aside the nanostrips have a twofold effect: they suppress strongly the Bean Livingston barrier \cite{willa} for the vortex entrance along the x-direction, such that $H_s \simeq H_{c1}$, and act as flux focusers for the self-generated magnetic field at the right and left edges of the strips. Both circumstances allow to neglect the motion of the magnetic field lines along the thickness of the sample ($z$-direction), so we assume that the vortex motion occurs exclusively along the $x$-direction.  We have estimated the critical current density of our samples in the presence of these effects. This has been accomplished by modelling the relation  between the magnetic field $\bf{H}$ and the current density $\bf{j}$ at the surface of the sample by using the Amp\`{e}re law and assuming that the critical current density is attained in correspondence of the four corner points. We get $j=4H/\alpha \gamma d$, and for the critical current density (see Supplementary Information):
\begin{equation}
\label{eq:jc1nano}
j_{c}=\frac{4 H_{c1}}{\alpha \gamma d}= \frac{ \phi_0}{  \pi \alpha \gamma \mu_0 \lambda(T)^2 d}\ln\left(\frac{\lambda}{\xi}\right),	
\end{equation}	
where $\alpha$ ($\alpha>1$) is a geometrical (demagnetizing) factor, taking into account both the sample geometry and the presence of the overlooking superconducting banks, $d$ is  the thickness of the strip. The magnetic field lines threads the nanostrip along the $x$ direction (see Fig. \ref{fig:micro}). The quantity $\gamma$ ($\gamma>1$) is a further magnetic field amplification factor attained at each one of the four corners defining the nanostrip. By using in equations (\ref{eq:Hc1}), the values $\lambda=560$ nm and $\xi=2$ nm, we obtain $B_{c1}(0)=29.6$ G. The $j_c$ value provided by equation (\ref{eq:jc1nano}) should be compared  with the value $j_c(0)=1.5 \times 10^5$A/cm$^2$ extrapolated from the measurements. This comparison gives for the quantity $\alpha \gamma$ an extrapolated value of $\sim 63$. Both $\alpha$ and $\gamma$ are hardly calculated \textit{a priori} in our samples. However by assuming $\gamma =(2/3)(w/\pi \xi)^{1/3}$ as estimated in ref \cite{clem},  we obtain $\gamma$, for sample A and sample B respectively, as $\gamma_A= 2.86$ and $\gamma_B=3.35$. We also obtain for the $\alpha$ parameters $\alpha_A=2\times 1.78$ and $\alpha_B=2 \times 2.02$ (see Supplementary Information). In this way a priori estimated values of $\alpha \gamma$, $\alpha \gamma \sim 10$ and  $\alpha \gamma \sim 13$, are obtained, which roughly approach the value extrapolated by the measurements.

\subsection*{Nanostrip single vortex creep flow equations, pinning potential determination, creep flow parameters}
\begin{figure}[htbp]
\centering
\includegraphics[width=15cm]{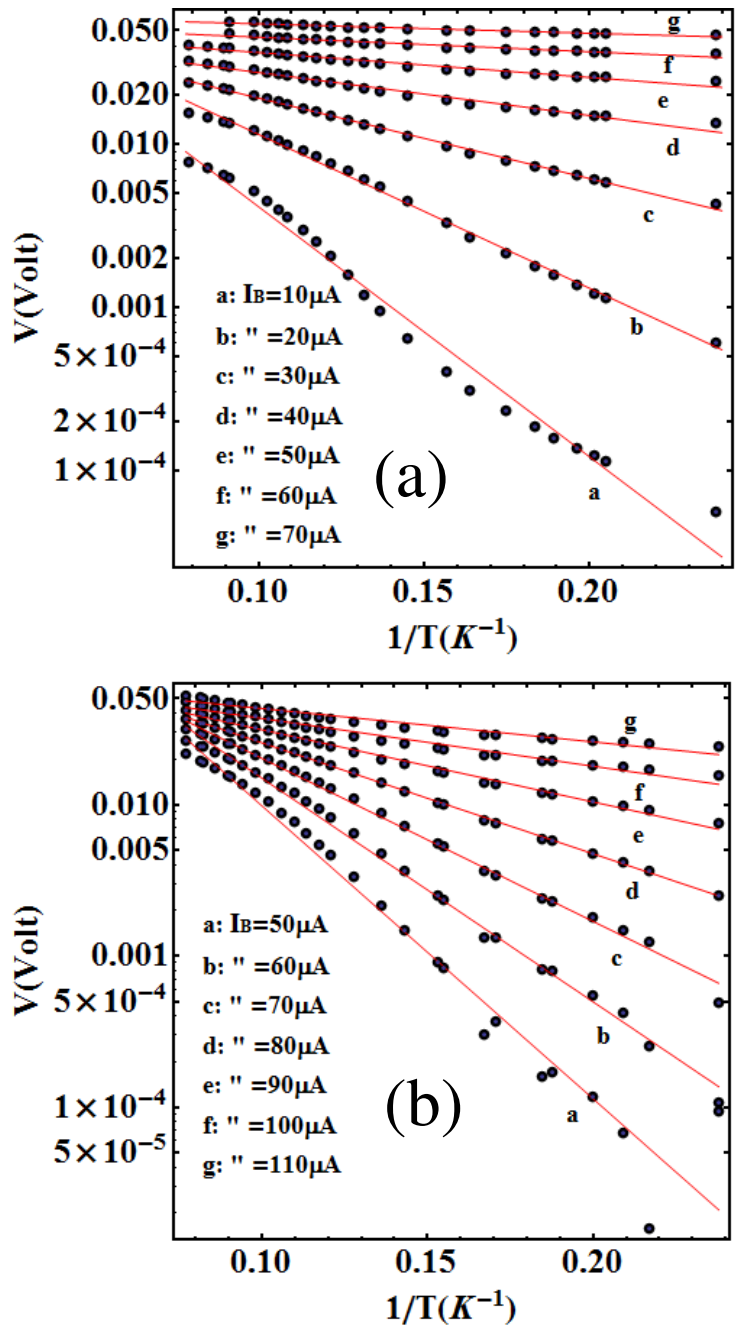}
 \caption{Inverse of the temperature dependence of the voltage when the sample ((a), sample A and (b), sample B)) is current-biased ($\log(V)-1/T$ relation) as obtained from data in Fig. \ref{fig:CVC}. Almost linear dependence of log(V) on 1/T is observed consistently  with the theoretical flux creep model, equation (\ref{eq:enpuku}) (red lines).}\label{fig:log}
\end{figure}

\begin{figure}[htbp]
\centering
\includegraphics[width=\linewidth]{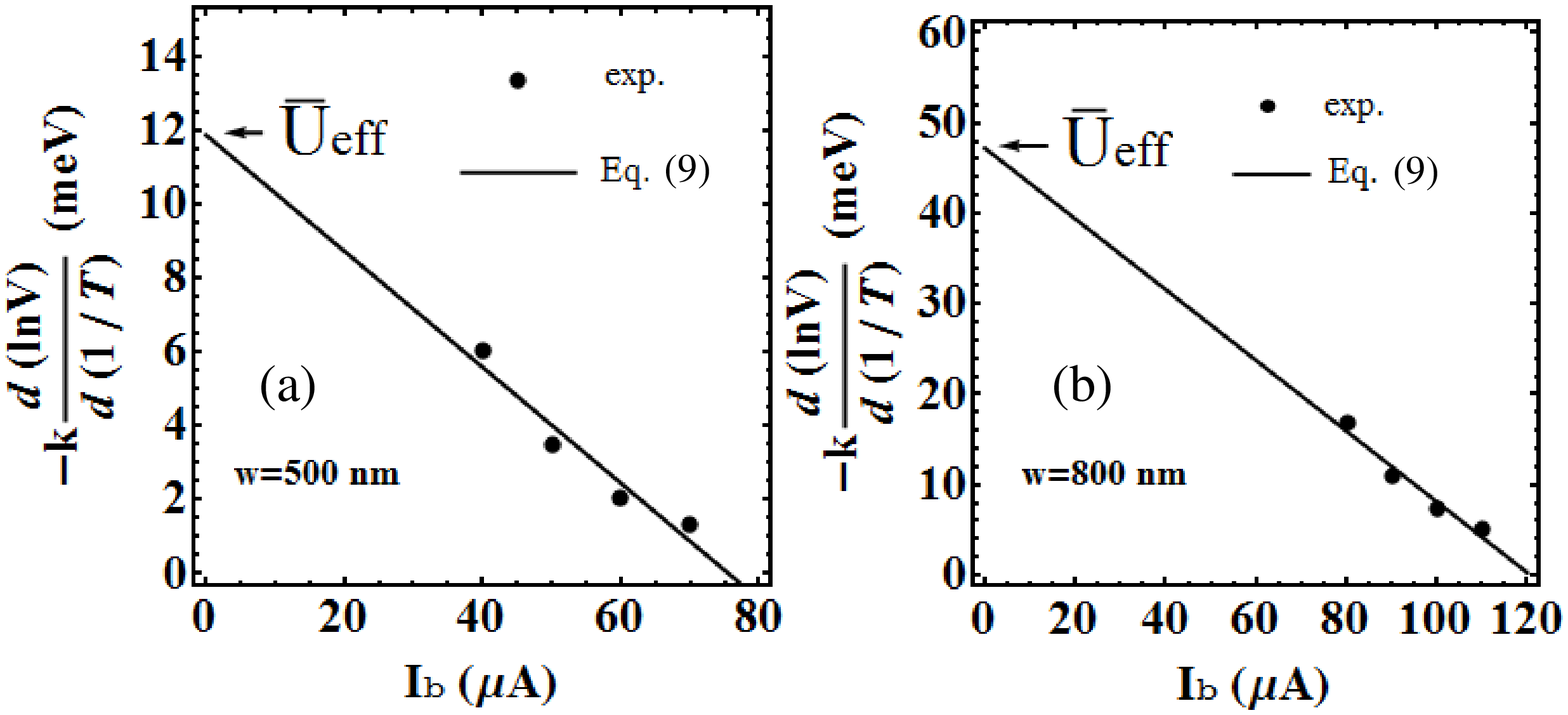}
 \caption{Experimental relation between $d(\ln V)/d(1/T)$ and the bias current $I_b$, for $T \ll T_c$ for sample A, (a), and sample B, (b). The value of $d(\ln V)/d(1/T)$ is obtained from the slope of the $\log(V)-1/T$ relation shown in Fig. \ref{fig:log} (a) and  Fig. \ref{fig:log} (b), respectively. We consider the lowest four temperatures and the corresponding four experimental values of $d(\ln V)/d(1/T)$. The solid line is equation (\ref{eq:minuslog1}) with $I_c(0)= 76$ $\mu$A, $U(0)=11.9$ meV for sample A, and $I_c(0)= 120$ $\mu$A, $U(0)=47.2$ meV for sample B. The two values of the $I_c(0)$ were  extrapolated from the experimental data illustrated in Fig. \ref{fig:linear}}\label{fig:sei}
\end{figure}

\begin{figure}[htbp]
\centering
\includegraphics[width=\linewidth]{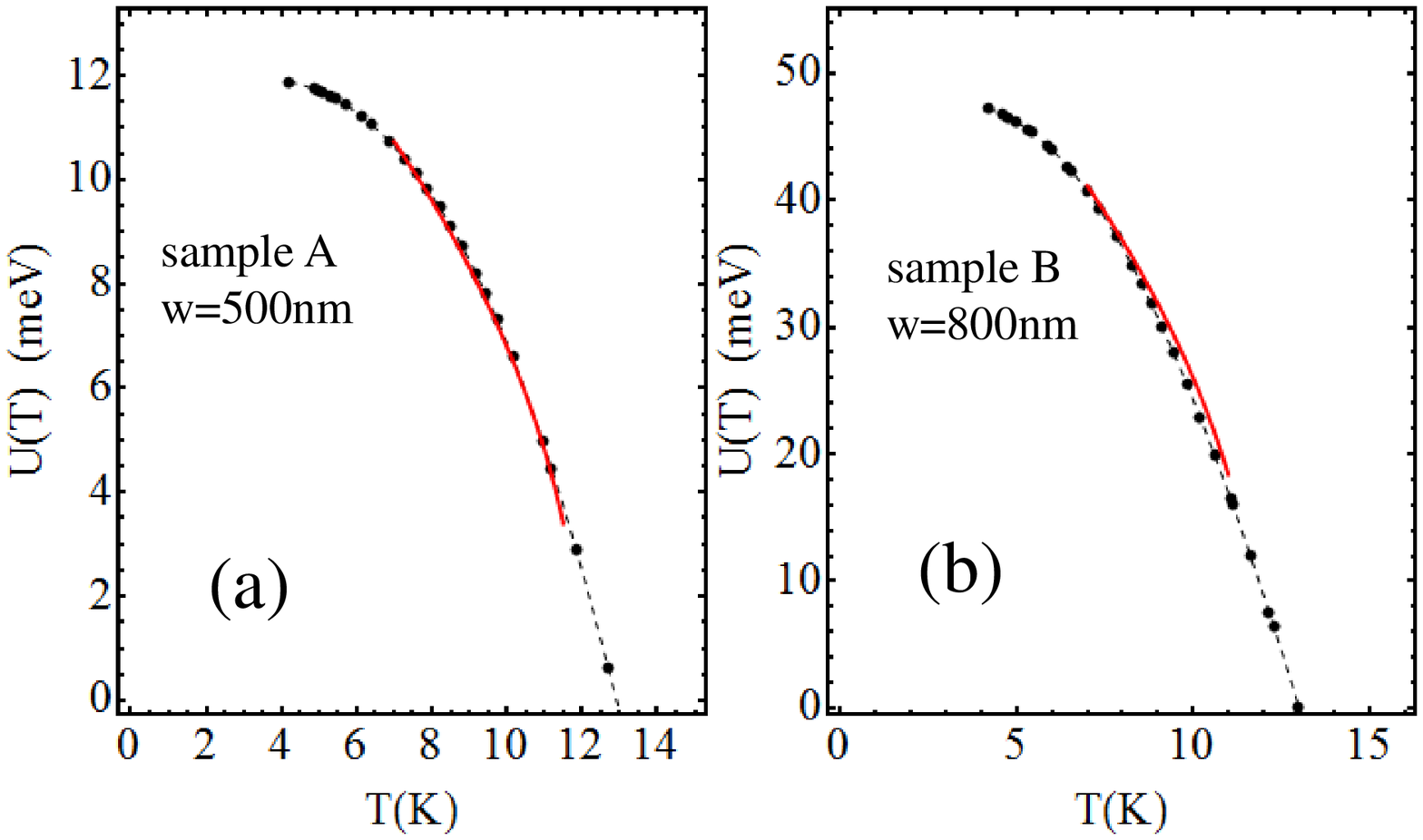}
 \caption{ Experimental result of the temperature dependence of the pinning potential U(T) for sample A, a),  and B, b).   The red line in a) and b) shows the relation $U(T)=1.4 U(0)\sqrt{(T_c-T)/T_c)}$, ($T_c$ = 12 K)}\label{fig:UT}
\end{figure}

Now we briefly derive the equations which describe the physics underlying the observed CVCs and allow a determination of the pinning energy. These equations are based on the Kim Anderson theory of creep flow \cite{kimanderson}. To be definite, we assume that the magnetic field lines penetrate in correspondence of the four corners defining the stripline and occupy two channels of area $w \times 2\lambda$ (see Fig. \ref{fig:micro} b). A train of $N$ magnetic flux lines (or two trains of opposite sign magnetic flux lines travelling half strip width, for channel) moving across the entire strip in the $x$-direction  induces a voltage $V$ at the strip terminals (see Fig. \ref{fig:micro}), which can be be written as
\begin{equation}
\label{eq:V}
V=Nv \phi_0/w
\end{equation}
where $v$ is the average velocity of a flux line crossing the strip. The number of vortices $N$ present in the channel depends on the intensity of the magnetic induction at the two edges of the channel, this latter depends, in turn, on the bias current $I_b$. In fact the average magnetic induction in the channel is $B=N\phi_0/2\lambda w$. On the other hand,  as shown in the previous section (See also Supplementary Information), at the edge,  $B=\mu_0 \alpha \gamma  I_b/\pi w$. By comparing the two expressions we obtain $N=2\mu_0 \alpha \gamma \lambda I_b/\pi \phi_0$. Since in our experiments $\alpha \gamma \sim 63$,   $I_b\sim 30\mu$A and $I_b\sim 120\mu$A at $V=5$mV,  respectively in sample A and B, we obtain the nominal value $N=2\mu_0 \alpha \gamma \lambda I_b/\pi \phi_0 \sim 0.4$ in sample A (that is, in average, one vortex (N=1) travelling the entire width or a vortex and an anti-vortex travelling half width and annihilating at the centre) and $N=2\mu_0 \alpha \gamma \lambda I_b/\pi \phi_0 \sim 1.6$ in sample B (that is, in average, two vortices (N=2) travelling the entire width or two vortices and two anti-vortices travelling half width and annihilating at the centre). Furthermore, from equation (\ref{eq:V}), supposing the presence of two channels generating the observed voltage of 5mV we obtain the velocities $v_A\sim 6.2 \times 10^5$m/s and $v_B\sim 5 \times 10^5$m/s. The realization of such  low-density vortex states has been predicted in ref. \cite{willa}  where a setup similar to the nanostrip used in the present work is studied and put in correlation with the suppression of the Bean Livingston barrier.  Substituting $N$, equation (\ref{eq:V}) becomes
\begin{equation}
\label{eq:V2}
V=\left(\frac{\mu_0 \alpha \gamma \lambda v}{\pi w}\right) I_b
\end{equation}
Taking into account the vortex thermal hopping mechanism and  neglecting backward hopping, i.e assuming ($W/k_BT\gtrsim 1$)\cite{dew}, the mean velocity in equation (\ref{eq:V}) may be written as
\begin{equation}
\label{eq:velocity}
v=v_{0}\exp\left(-\frac{U}{k_BT}\right)\exp\left(\frac{W}{k_BT}\right)	
\end{equation}
(see Supplementary Information) where $v_{0}$ is the maximum vortex creep velocity ($v<v_0$), $U$ is the pinning potential at temperature $T$, $W$ is the work done by the mean Lorentz force $j_b d \phi_0$  during the motion of one vortex from a pinning site to the other. The velocity  $v_{0}$ may be written in terms of  $\delta$, the  effective  pinning potential range, and $\omega_0$, the attempt frequency for a magnetic flux line to hop over an energy barrier $U$ and move on a distance $\delta$ 
 \begin{equation}
\label{eq:velocity0}
v_{0}= \omega_0	\delta.
\end{equation}
Assuming $\delta \sim 10 \xi$, attempt  frequencies in the range $10^{13}$Hz (as found in (Y-Ba-Cu-O) \cite{pann}) are required to realize velocities of the order of $10^5$m/s found above. The pinning sites can be described as potential wells and the work $W$ can be written as $W \sim (j_b d\phi_0)\delta= I_b\phi_0\delta/w$. 
Then the voltage equation (\ref{eq:V}) gives 
\begin{eqnarray}
\label{eq:toenpuku}
&&V=I_b\left(\frac{2\mu_0 \alpha \gamma \lambda \omega_0 \delta}{\pi w}\right) \exp \left(-\frac{\Delta U}{k_B T}\right),\nonumber \\
&&\Delta U(T,I_b) =U(T) - I_b\phi_0\delta/w.	
\end{eqnarray}
where $\Delta U$ is the energy barrier against creep flow. $U(T)$ represents the temperature dependent pinning potential.  Now we observe that the case $\Delta U \sim 0$ corresponds to a bias current $I_b=I_c(T)$ implying that $U(T)= I_{c}(T)\phi_0 \delta/w$. Thus at $T=0$ results  $\delta = U(0)w/I_c(0)\phi_0$. Then equation ($\ref{eq:toenpuku}$) writes \cite{enpu1}
\begin{eqnarray}
\label{eq:enpuku}
&&V=A  \exp(-\Delta U/k_BT), \nonumber \\			
&&\Delta U=U(T) - I_b U(0)/ I_c(0),			
\end{eqnarray}			
where $ A=2I_b{ \mu_0 \alpha \gamma \omega_0 \lambda U(0)}/{ \pi \phi_0 I_c(0)}$ is a constant independent from the temperature.
Equations (\ref{eq:enpuku}), a single vortex creep flow model, have been  used in the past by Enpuku $\textit{et al.}$ \cite{enpu1,enpu2}  to establish YBCO thin film properties. In the Enpuku method, the pinning potential is determined by measuring the temperature dependence of the CVC. 
 The experimental values of $\log(V)$ as a function of the inverse of the temperature $1/T$ are considered to obtain the pinning potential $U$ by means of equation (\ref{eq:enpuku}). 
 In Fig. \ref{fig:log} we show experimental results of the $\log(V)-1/T$ relation when sample A and sample B are current biased for the values indicated in the legend. As can be seen, the value of $\log(V)$ decreases linearly with 1/T, which is consistent with the creep flow interpretation and the theoretical predictions of equation (\ref{eq:toenpuku}) or (\ref{eq:enpuku}).
Firstly,  when $ T<<T_c$, we can assume $U(T)\sim U(0)$,  and  $V=A\exp(-\Delta U/k_BT)$ with $\Delta U=U(0)(1 - I_b / I_c(0))$,  so that equation (\ref{eq:enpuku}) becomes $V=A \exp[-U(0)( 1 - I_b / I_c(0))/k_BT]	$.						
 The negative slope value of the $\log(V)-1/T$ dependence can be related to the effective potential energy $U_{eff}$ through the expression
\begin{equation}
\label{eq:minuslog1}
U_{eff}=-k_B\frac{d(\ln(V))}{d(1/T)}=
U(0)\left(1- \frac{I_b}{I_c(0)}\right), \hspace{1.0cm} \textrm{for}  \hspace{0.3cm}T<<T_c,
\end{equation}
so that $U(0)=\bar{U}_{eff}=U_{eff}|_{I_b=0}$. In Fig. \ref{fig:sei} the experimental value of $-k_Bd(\ln(V))/d(1/T)$ is shown as a function of $I_b$ as obtained by the experimental data of Fig. \ref{fig:log} for the two samples A and B considering the four lowest temperatures. In agreement with equation (\ref{eq:minuslog1}), the value of $-k_Bd(\ln(V))/d(1/T)$ decreases approximately linearly with $I_b$, supporting the hypothesis that the flux creep dominates the CVC in the low bias current regime.
From the comparison between the experimental results and equation (\ref{eq:minuslog1}), illustrated in Fig. \ref{fig:sei},  we obtain the pinning potential  $U(0)=\bar{U}_{eff}=11.9$ meV for sample A, and $U(0)=\bar{U}_{eff}=47.2$ meV for sample B. In carrying out this comparison we used the values of $I_c(0)$ obtained by the extrapolation of the $I_c-T$ relations shown in Fig. \ref{fig:linear}, that is,  $I_c(0)=76$ $\mu A$ ($j_c(0)=1.52 \times 10^5$ A/cm$^2$) for sample A  and $I_c(0)=120$ $\mu$A ($j_c(0)=1.50 \times 10^5$ A/cm$^2$) for sample B.
As can be seen, and rather unexpectedly, different values of the energy $U$ are obtained for the two samples considered, i.e.  $12$meV and $47$meV respectively. We argue that this difference between the two samples is in relation with the very small number of vortices involved in the creep flow, rather than correlated to the two different sample widths, 500nm and 800nm. The explored pinning sites are limited in number so that the pinning energy returned characterizes the particular landscape experienced by the few vortices in a limited portion of the two samples. In larger samples there are more vortices more uniformly distributed,  such that the variance of potentially determined value of $U(0)$ would be lower. We conclude this section by noting that the  dependence  from the temperature  of the pinning energy $U(T)$ can be also experimentally derived by using
equation (\ref{eq:enpuku}) and dropping the condition $T\ll T_c$. One obtains \cite{enpu1}:
 \begin{equation}
 \label{eq:enpukuUT}
 U(T)=U(T_1)\frac{T}{T_1}+T\int_{T_1}^T\frac{1}{T^2}\left[k_B\frac{d(\log V)}{d(1/T)}-U(0)\frac{I_b}{I_c(0)}\right]dT
 \end{equation}
where $T_1$ is an integration constant ($T_1=4.2$K). In equation (\ref{eq:enpukuUT}) the value of $d(\log V)/d(1/T)$ as a function of the temperature can be experimentally obtained from the $\log(V) - 1/T$ relation shown in Fig.\ref{fig:log}. Once the integrand has been evaluated in this way, by using the values of $U(0)$ and $I_c(0)$  previous found, and performing the integration in equation (\ref{eq:enpukuUT}) numerically, we obtain the temperature dependence $U(T)$. Figure \ref{fig:UT}, a) and b),  shows the result of this procedure for sample A and sample B respectively. The red line on the same figure, is a fit with the Ginzburg Landau theory as explained further into the text. 

\subsection*{Correction for nanostrips}
In nanostrips with w$< \lambda$, the simultaneous presence  of  magnetic flux lines of opposite sign enucleated at opposite edges within a distance $ l<\lambda$ must be taken into account. Indeed there is an extra force, acting all the time and independently from the bias current,  which contributes to push inward couples of magnetic flux lines with opposite sign attracting each other. An additional energy contribution $W_f$  adds  to the work $W \sim (j_b d\phi_0)\delta$ done by the average Lorentz force and can be roughly estimated as follows. The attractive force per unit length between two vortices of opposite sign separated by a distance $l$ is given by \cite{orl}
\begin{equation}
\label{eq:force}
 f=\frac{\phi_0^2}{2 \pi \mu_0 \lambda^3}K_1\left(\frac{l}{\lambda}\right),
\end{equation}
where $K_1$ is the modified Bessel function of second kind of order one. We assume that the two magnetic flux lines from opposite edges hop between pinning sites in steps of length $\delta$ towards the centre of the bridge where they self-annihilate. The separation distance of the two flux lines ranges between a maximum $l_m \sim w$ and a minimum distance before annihilation which is approximately equal to $\xi$. The work done by the mean attractive force per unit length $\bar{f}$ ($\bar{f}=\int_{\xi}^{l_m} f dl/(l_m-\xi)$) during the motion of one vortex from one pinning site to another is $W_f=\bar{f} \delta d$, that is
\begin{equation}
\label{eq:work}
 W_f=\delta d \frac{\phi_0^2}{2 \pi \mu_0 \lambda^2}\frac{K_0\left(\frac{\xi}{\lambda}\right)-K_0\left(\frac{l_m}{\lambda}\right)}{l_m-\xi}\sim \delta d \frac{\phi_0^2}{2 \pi \mu_0 \lambda^2} \frac {K_0\left(\frac{\xi}{\lambda}\right)}{l_m}\sim \alpha \delta \gamma j_{c}\phi_0 \frac{d^2}{2w} .
\end{equation}
where we have used equation (\ref{eq:jc1nano}) and the approximation $K_0(\xi/\lambda)\sim \ln(\lambda/\xi)$, valid for $\lambda \gg\ \xi$.
Note that in a large width bridge ($w \gg d$) this energy contribution is small and can be neglected. We now evaluate the effect on $U(0)$ of the correction due to $W_f$. The barrier against creep with the introduction of $W_f=\bar{f}\delta d$ becomes
\begin{equation}
\Delta U(T,I_b)=U(T) -j_b \phi_0 \delta d-W_f.
\end{equation}
The critical condition for the suppression of the barrier against creep, i. e.  $\Delta U\sim 0$, occurring at $j_b=j_{c}$, gives
\begin{equation}
\label{eq:delta}
\delta=\frac{U(T)}{j_c \phi_0 d\left(1+\frac{\alpha \gamma d}{2w}\right) }.
\end{equation}
Equation (\ref{eq:enpuku}) generalizes to
\begin{eqnarray}
\label{eq:enpukurivisited}
&&V=A  \exp(-\Delta U/k_BT) \nonumber \\					
&&\Delta U=U(T) - U(0)\frac{j_b+j_c\frac{\alpha \gamma d}{2w}}{j_c  \left(1+\frac{\alpha \gamma d}{2 w}\right)};		
\end{eqnarray}	
where we have used  equation (\ref{eq:delta}) at $T=0$. For $T \ll T_c$, we assume $U(T) \sim U(0)$ and we have
\begin{equation}
\label{eq:minuslog}
U_{eff}=-k\frac{d(\ln(V))}{d(1/T)}=
U(0)\left(1+\frac{\alpha \gamma d}{2 w}\right)^{-1}\left(1-\frac{I_b}{I_c(0)}\right).\\ \nonumber	
\end{equation}
The comparison with the experimental data provides now for $U(0)$
\begin{equation}
\label{eq:U(0)}
U(0)=\bar{U}_{eff}\left( 1+\frac{\alpha \gamma d }{2 w}\right)
\end{equation}
where again $\bar{U}_{eff}=U_{eff}|_{I_b=0}$.  As summarized in Table \ref{tab:table2}, in which $\alpha \gamma \sim 63$, the obtained values of  $U(0)$, $\sim 87$ meV and $\sim 233$ meV for sample A and sample B, respectively, are larger than the estimations done by using the basic Enpuku \textit{et al.} model, i.e. 11.9 and 47.2 meV.  Through equation (\ref{eq:delta}) it is possible to estimate also the pinning potential range $\delta(0)$, that results $\delta \sim 6$ nm and $\delta \sim 25$ nm for sample A and B, respectively. 
\begin{table}[ht]
\centering
\begin{tabular}{cccccccc}
Sample&\mbox{$w$(nm)}&\mbox{$\left(1+\alpha \gamma d /  2w\right)$}&\mbox{$I_c(0)$($ \mu A$)}&\mbox{$j_c(0)$(A/cm$^2$)} &\mbox{$\bar{U}_{eff}$(meV)}&\mbox{$U(0)$(meV)}&\mbox{$\delta$(nm)}\\
\hline
A&\mbox{500}&\mbox{7.3}&\mbox{76}&\mbox{$1.52 \times 10^5$ }&\mbox{11.9}&\mbox{87}&\mbox{6}\\
B&\mbox{800}&\mbox{4.94}&\mbox{120}&\mbox{$1.50 \times 10^5$ }&\mbox{47.2}&\mbox{233}&\mbox{25}\\
\end{tabular}
\caption{\label{tab:table2} Parameters of nanostrips, U(0) is the pinning potential at T=0, corrected for $W_f$ (Equation (\ref{eq:U(0)}))}
\end{table}
\subsection*{Discriminating defect types}
It is interesting to compare the found $U(0)$ values with the theoretical estimates obtained by relating the pinning energy with the kind of defect. These estimates are based on the general consideration that the condensation energy   $V_c \mu_0 H_c^2/2$  of a volume $V_c$ of the vortex core can be saved if the flux line core passes through a region where the order parameter is already zero. Here $H_c$ is the thermodynamical critical field  given by $H_c=H_{c2}/\kappa \sqrt{2} = \phi_0^2/(\kappa \sqrt{2^3} \pi \mu_0 \xi^2)$, where $\kappa=\lambda/\xi$ and $H_{c2}$ is the upper critical magnetic field.  For point like defects consisting of a small spherical void ($V_c \sim 4 \pi/3 (D/2)^3$ ), or a non superconductive inclusion, of diameter $D$, smaller than the coherence length $\xi$,  $U$  is given by \cite{ullm}
\begin{equation}
\label{eq:smallvoid}
U=\frac{(2 \pi)^{3/2}B_{c2}^{5/2}}{48 \phi_0^{1/2}\kappa^2 \mu_0}\xi D^3
\end{equation}
Even with $D = \xi$, one obtains a pinning energy as small as $U = 0.732$ meV with $B_{c2}=100$ T, $\kappa=560$ nm$/2$ nm (corresponding to a thermodynamic field $\mu_0 H_c$ of 0.25 T). A high number (order of hundreds) of small point-like type defects is expected to  pin the magnetic flux line through the thickness $d$ of the nanostrips.
For a void larger than the core region  the maximum pinning energy depends on the shape and orientation of the void. For the  case of a sharp void surface ($V_c \sim \pi \xi^2 L_z$) of length $L_z $ (i.e. occupying all the thickness) parallel to the vortex, the pinning energy is given by \cite{ullm}			
\begin{equation}
\label{eq:sharpsurface}
U= (2 \pi \phi_0)^{1/2} \frac {B_{c2}^{3/2}}{\kappa^2 \mu_0}\xi L_z
\end{equation}

which results in $U(0) = 1446$ meV for a void occupying the whole thickness of the film  $L_z=d$ and with $B_{c2}=100$ T. This suggest  2 D defects extending over a large fraction  of the thickness. Assuming the temperature dependencies of $H_c$ and $\xi$ given by the Ginzburg-Landau theory, we obtain from equation (\ref{eq:smallvoid}) the temperature dependence of U as $U(T)/U(0)=\eta(1-T/T_c)^{1/2}$, where $\eta$ is a parameter close to one. In Fig. \ref{fig:UT} the solid line shows the theoretical result with $\eta=1.4$; the critical temperature has been chosen as $T_c=12$ K (instead of $13$ K) so as to fit the theoretical values to the experimental values. As can be seen, the experimental temperature dependence of $U$ is satisfactorily reproduced. Work is in progress now to study the detailed nature of $U$. Besides conventional mechanisms due to defects of the material,  pinning originated by the interaction of fluxons with the local magnetization could be also considered in Fe(Se,Te). Also in this case, thermally activation mechanism of the self-generated flux lines and uncorrelated motion of the flux lines can explain our experimental data. 

\subsection*{Conclusions}
In summary, we have studied the resistive state induced by the current in Fe(Se$_{0.5}$,Te$_{0.5}$) superconducting nanostrips (width $w$ less than the London length $\lambda$), in view of a potential application of iron based superconductors in the field of electronics and nano-electronics. The resistive state emerging at low currents in the collected  CVCs of the two samples is  due to the depinning (creep flow) of a very limited number of magnetic field lines. To make progress we use a creep flow model used in the past to characterize YBCO strip. The pinning potential values of few tens of meV, provided as output of the model, are low in comparison with those found typically in literature. We individuate in the attraction between vortices of opposite signs coming from the two edges of the strips the mechanism to introduce into in order to extend applicability of the model to the nanostrip case and restore agreement. Two important points result also from our analysis of FIB (Fast Ion Bombardment) fabricated nanostrips: the evidence of a reduced Bean Livingston barrier caused by the presence of superconducting banks aside the nanostrips and the overwhelming role of sharp corners driving the entrance of magnetic field lines in the nanostrips. Taking into account these aspects, the conventional model of creep flow allows a suitable description of the transport properties also in the case considered of very narrow nanometric striplines.

\section*{Methods}
\subsection*{Fabrication and measurement setup}
Stoichiometric fluctuations in the films were quantified with a computational approach from scanning tunnelling microscopy images. A nominal stoichiometry FeSe$0.45\pm0.06$Te$0.55\pm0.09$ is estimated \cite{pera}. Indeed, considering only the error values, one might approximately conclude that the chalcogenides concentration is compatible with an Se/Te stoichiometry of 50 and 50 per cent.
Fe(Se$_{0.5}$,Te$_{0.5}$) highly oriented thin films base electrodes were
prepared by laser ablation. A Nd:YAG laser beam at 1024 nm with  2 mm$^2$ spot area and fluency 0.5 J cm$^{-2}$ is focused on the target at a repetition rate of 3 Hz. The target is positioned at 5 cm of distance from the beam. Fe(Se$_0.5$,Te$_0.5$) films 100 nm thick are deposited on CaF$_2$ single crystal substrates. Further details may be found in refs.  \cite{bell1,bell2}. The patterning of the nanostrips has been done through two different steps. First, micrometric strips are defined by standard photolithography and ion milling etching, then the stripline dimensions are further reduced by FIB (Focused Ion Beam) using Ga-ions. For the measurements of the current-voltage characteristics the films were patterned in nanostrips $L=3$ $\mu$m in length and width $w=500$ nm (sample A) and $w=800$ nm (sample B). $S_A=5 \times 10^{-14}$ m$^2$ and $S_B=8 \times 10^{-14}$ m$^2$ are the nominal cross section surfaces. $A_A=1.5 \times 10^{-12}$ m$^2$ $A_B=2.4 \times 10^{-12}$ m$^2$. The critical current density of the patterned strips at $T=4.2$ K was $j_c=3.2 \times 10^4$ A/cm$^2$ (sample A) and $j_c=8.1 \times 10^5$A/cm$^2$ (sample B) (\ref{tab:table1}). Current voltage characteristics have been collected in a temperature interval ranging from 4.2 K to 13 K (Sample A, 24 curves, Sample B, 27 curves) (see Fig.\ref{fig:CVC}). Here $T_c$ of the nanostrip is defined as the highest temperature to which the derivative $dI_b/dV$ shows a peak at $V=0$. Above this temperature the CVCs show an ohmic behaviour and, as a consequence, the peak in the $dI_b/dV$ disappears. We explicitly notice that the measured curves, when considered in the full range of  currents (not shown in Fig. \ref{fig:CVC}) extrapolate to zero current. During measurements, samples are located under vacuum inside a cryogenic probe. The temperature exchange with the thermal bath is obtained with 0.13 mbar helium gas. The  cryogenic insert is shielded by two small concentric lead (internal) and cryoperm (external) cylinders, both at 4.2 K. Samples are measured in a highly shielded cryostat, surrounded by three $\mu$-metal cylinders and one external aluminum shield, with an attenuation factor $S > 10^4$.  Four-point contact method has been used for measuring nanostrips CVCs. Wires are filtered through low-pass passive filters, and electronics is powered with dc batteries. Samples are controlled through DAC/ADC board and data are directly collected to a PC. No external magnetic field was applied during the measurement and the presence of suitable magnetic shields granted that the superconductive transition did not originate magnetic flux trapping \cite{stan}.

\section*{Acknowledgements}

This work has been partially supported by the financial
contribution of MIUR, Progetto premiale 2015 "Q-SecGroundSpace" and EU NMP.2011.2.2-6 IRONSEA Project
No. 283141. FIB nanostructurations and SEM inspections have been performed at Nanofacility  Piemonte INRIM, a laboratory supported by Compagnia di San Paolo Foundation. 

\section*{Author contributions statement}

E.S. designed the project, E. S. and C.C. performed the experiments, E.E. made the nanopatterning process, C.N. wrote the paper and made the simulations for the data analysis.  E. B., V.B., and C.F.  provided the thin film samples. All authors contributed to the concept and revised the manuscript. 

\section*{Additional information}
The authors declare no competing financial interests.


\begin{thebibliography}{apssamp}
\bibitem{note} The voltage threshold is fixed to a value reasonably above the voltage  noise level such to limit as much as possible scattering of data. In our case 130$\mu$V.
\end{thebibliography}
\end{document}